\documentclass[twocolumn,secnumarabic,bibnotes,aps,prl, floatfix]{revtex4-1}
\usepackage{amssymb,amsmath,amsthm,bm}
\usepackage{graphicx}
\usepackage[breaklinks=true,colorlinks=true,linkcolor=blue,urlcolor=blue,citecolor=blue]{hyperref}

\begin{document}

\title{Scaling Analysis of Anomalous Hall Resistivity in the Co$_{2}$TiAl Heusler Alloy}%

\author{Rudra Prasad Jena}
\author{Devendra Kumar}
\author{Archana Lakhani}
\email{archnalakhani@gmail.com}
\affiliation{UGC-DAE Consortium for Scientific Research, Devi Ahilya University Campus, Khandwa Road, Indore-452001, M.P., India}

\begin{abstract}
	A comprehensive magnetotransport study including resistivity ($\rho_{xx}$) at various fields, isothermal magnetoresistance and Hall resistivity ($\rho_{xy}$) has been carried out at different temperatures on the Co$_{2}$TiAl Heusler alloy. Co$_{2}$TiAl alloy shows a paramagnetic (PM) to ferromagnetic (FM) transition below the curie temperature (T$_{C}$) $\sim$ 125 K. In the FM region, resistivity and magnetoresistance reveals a spin flip electron-magnon scattering and the Hall resistivity unveils the anomalous Hall resistivity ($\rho_{xy}^{AH}$).  Scaling of anomalous Hall resistivity with resistivity establishes the extrinsic scattering process responsible for the anomalous hall resistivity; however Skew scattering is the dominant mechanism compared to the side-jump contribution. A one to one correspondence between magnetoresistance and side-jump contribution to anomalous Hall resistivity verifies the electron-magnon scattering being the source of side-jump contribution to the anomalous hall resistivity.
\end{abstract}
\maketitle

\section{Introduction}
Ferromagnetic Heusler alloys (HA) are known for their multi-functional properties such as shape memory effect, magneto-caloric effect, large magnetic field induced strain, high spin polarization, topological properties, large anomalous hall effect etc. \cite{felser2015book,graf2013handbook,heusler4.0,ludbrook2017sr}. In recent times, the Cobalt based Heusler alloys have attracted the attention of researchers due to functional properties like half metallicity, spin polarization and anomalous hall effect (AHE) which makes them possible candidates for spintronics applications \cite{sakuraba2005,jourdan2014,bainsla2016apr,varaprasad2012acta}.

Anomalous Hall effect is observed in the ferromagnetic materials resulting from spontaneous magnetization of the material hence present even in absence of magnetic field and can be described by the equation:
\begin{equation}\label{ahe eqn}
	\rho_{xy} = \rho_{xy}^{OH} + \rho_{xy}^{AH} = R_{0}H + 4\pi M_{s}R_{s}
\end{equation}
where $ \rho_{xy}^{OH} $ and $\rho_{xy}^{AH}$ are the ordinary and anomalous hall resistivity respectively, and $H$ is the applied magnetic field. The coefficients $R_{0}$ and $R_{S}$ are characterized by the strength of ordinary ($ \rho_{xy}^{OH} $) and anomalous ($ \rho_{xy}^{AH} $) Hall resistivity \cite{nagaosa2010}. The ordinary Hall effect (OHE) is classically explained by the Lorentz force deflecting the moving charge carriers whereas, for anomalous hall effect, in general, three scattering mechanisms are considered in the literature that explain the origin of anomalous hall resistivity. One of them is referred to as Smit asymmetric scattering or skew scattering mechanism \cite{smit1955,smit1958}, originally derived for scattering by impurities, for which anomalous hall resistivity is linearly proportional to longitudinal resistivity, $ \rho_{xy}^{AH-SK} \propto \rho_{xx}$. The second one is side-jump mechanism proposed by Berger \cite{berger1970prb}, which follows a quadratic dependence with longitudinal resistivity, $\rho_{xy}^{AH-SJ} \propto \rho_{xx}^2$. Both of these mechanisms have extrinsic origin, while the third mechanism is the intrinsic one which arises from spin orbit coupling and depends on the band structure inherent to the material, leading to $ \rho_{xy}^{AH-I} \propto \rho_{xx}^2 $. This was discovered by Karplus and Luttinger \cite{KL1954,luttinger1958}. Recently, Karplus Luttinger (KL) mechanism has been interpreted in the language of Berry curvature formalism \cite{yao2004prl,xiao2010rmv,kubler2012prb}.

Hall effect has been studied in Cobalt based Heusler alloys also, in order to understand their scattering mechanism. For example, a temperature independent anomalous hall effect proportional to magnetization has been observed in Co$ _{2}$CrAl suggesting the intrinsic mechanism in this alloy \cite{husmann2006prb}. The AHE study on the thin films of Co$ _{2}$FeAl and Co$_{2}$FeSi by Imort et al. suggested the skew scattering mechanism influenced by crystalline quality \cite{imort2012jap}, whereas in another detailed study by Hazra et al. on Co$_{2}$FeSi thin films, side-jump and skew scattering are found to be the dominant mechanism in comparison to the intrinsic contribution to anomalous hall resistivity \cite{hazra2017prb}. For Co$_{2}$MnSi$_{1-x}$Al$_{x}$ alloys the AHE scaling with resistivity showed the contributions both from skew scattering as well as intrinsic mechanism \cite{prestigiacomo2014jap}. The phenomenon of anomalous hall effect is intriguing due to the rich physics, complexity involved in understanding its origin and the possibility of applications in electronic devices.
\par
Co$_{2}$TiAl is a cobalt based Heusler alloy having L2$_{1}$ crystal structure with space group Fm$\bar{3}$m. It undergoes a PM to FM second order phase transition at $\sim$125 K. Cobalt atom contributes to the magnetic moment of this alloy and Co$_{2}$TiAl shows the typical characteristic of soft ferromagnetic material with saturation magnetic moment $\sim$ 0.7 $\mu_{B}$ \cite{webster1973,graf2009} . In this paper, we report for the first time a comprehensive magnetotransport study on Co$_{2}$TiAl alloy including resistivity, magnetoresistance and Hall resistivity. Co$_{2}$TiAl is the least explored cobalt based Heusler alloy specially in case of transport studies, whereas there are few magnetization studies present in the literature \cite{zhang2007jalcom}. A detailed Magnetoresistance and AHE analysis explains the transport mechanism in this alloy.

\section{Experimental details}

Co$_{2}$TiAl is synthesized by arc melting the stoichiometric amounts of Co, Ti and Al in high purity argon atmosphere. The as-cast ingot obtained is vacuum annealed (better than 5 x 10$ ^{-6} $ mbar) at 1073 K for a week. The sample is characterized by powder X-ray diffraction (XRD) using the Bruker D8 Advance Diffractometer (Cu K$ _\alpha $ = 1.54 $\mbox{\AA}$) for which a part of the sample was crushed using agate mortar to obtain the powdered specimen. A rectangular bar shaped piece from the remaining sample was cut for transport measurements. Resistivity and magnetoresistance are measured by standard four probe technique whereas a five probe method is used for measuring the Hall resistivity using the AC-Transport option of PPMS down to 2 K and magnetic fields up to 9 T.
\section{Results and disscussion}

\subsection{X-ray diffraction}
\begin{figure}[h!]
	\begin{minipage}{0.6\linewidth}
		\centering
		\includegraphics[width=\linewidth]{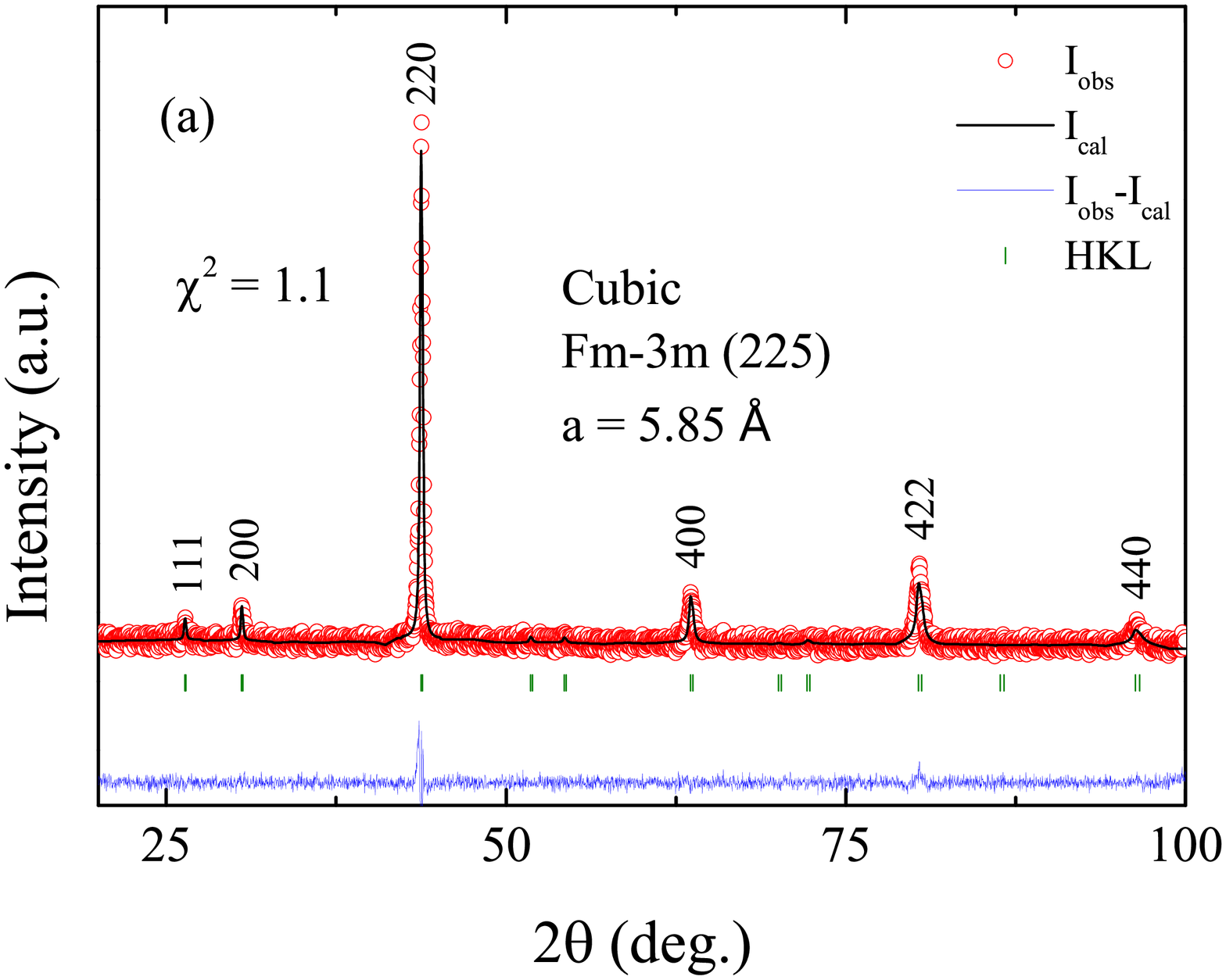}
	\end{minipage}
	\begin{minipage}{0.35\linewidth}
		\centering
		\includegraphics[width=\linewidth]{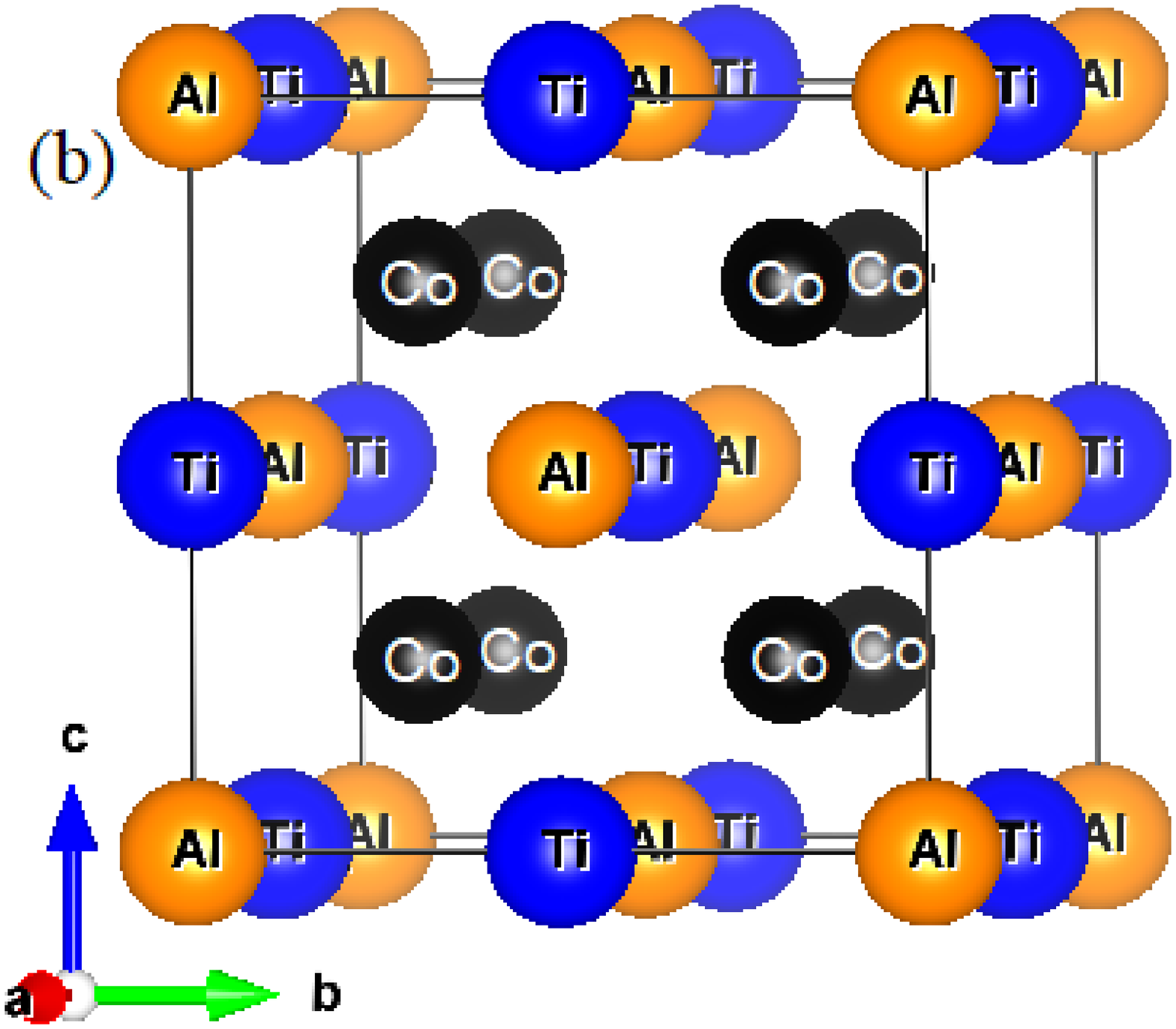}
	\end{minipage}
	\caption{(a) Red solid circles show the room temperature x-ray powder diffraction pattern along with its Rietveld refinement (black line). The blue line shows the difference between observed and calculated data along with the Bragg peaks (vertical green lines). (b) Crystal structure of Co$_{2}$TiAl.}
	\label{xrd-str}
\end{figure}
Figure \ref{xrd-str} (a) shows the x-ray powder diffraction pattern along with its Rietveld refinement, the difference between observed and fitted pattern along with the Bragg peak positions and figure \ref{xrd-str} (b)  shows the cubic crystal structure of the sample. The XRD pattern shows no impurity peaks. Rietveld refinement of the XRD data confirms the single phase nature of the sample and its crystallization in cubic L2$_{1}$ structure having space group Fm$\bar{3}$m. Cobalt occupies ($\frac{1}{4},\frac{1}{4},\frac{1}{4}$) whereas Ti and Al occupy (0,0,0) and ($\frac{1}{2},\frac{1}{2},\frac{1}{2}$) Wyckoff positions respectively. The lattice parameter obtained from Rietveld refinement is found to be 5.85$\mbox{\AA}$ which is in agreement with the literature \cite{webster1973}.The Rietveld refinement also suggests the stoichiometric formation of the sample i.e. Co:Ti:Al = 2.04(1):0.99(1):1.03(2).
\subsection{Resistivity and Magnetoresistance}

\begin{figure}[h!]
	\centering
	\includegraphics[width=0.7\linewidth]{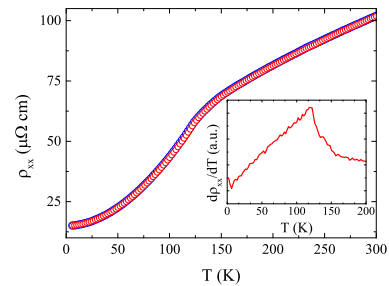}
	\caption{Resistivity versus temperature in zero magnetic field, blue and red circles indicate the cooling and heating curves. Inset shows the derivative of the resistivity data.}
	\label{RT_0T}
\end{figure}

Figure \ref{RT_0T} shows the temperature dependent resistivity in zero magnetic field, taken during cooling and heating cycles. Resistivity decreases gradually as the temperature is decreased from 300 K down to 5 K signifying the metallic nature of the sample in this temperature range.
\begin{figure}[b!]
	\centering
	\includegraphics[width=0.8\linewidth]{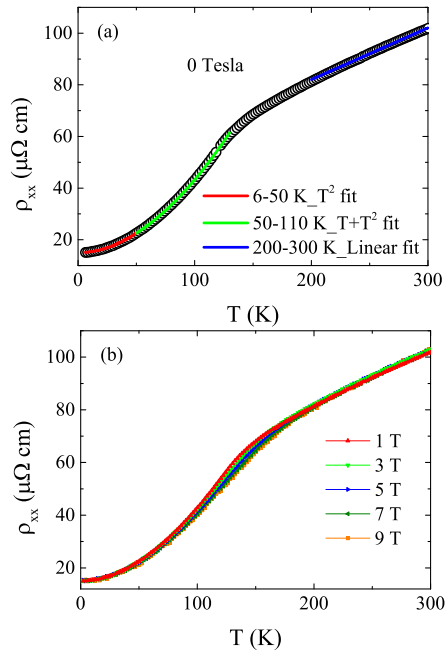}
	\caption{(a) Theoretical fits to the zero-field resistivity data as a function of temperature and (b) $\rho_{xx}$ (T) at various constant magnetic fields.}
	\label{rtfitall}
\end{figure}
Resistivity changes slope $\sim$ 125 K which corresponds to the temperature driven phase transition from PM to FM state at the Curie temperature (T$ _{C} $ = 125 K) \cite{graf2009}. T$ _{C} $ here is determined as the point of inflection in the derivative of zero field resistivity curve as shown in the inset of figure \ref{RT_0T}. The Residual Resistivity Ratio (RRR) value, defined as $\rho_{300K}/\rho_{5K}$, is 6.75 for the sample. According to Matthiessen's rule, the total resistivity of crystalline metallic samples is the sum of all the resistivity contributions resulting from various scattering processes which can be expressed as:
\begin{equation}\label{Rxx}
	\rho (T) = \rho_{0} + \rho_{e-m} + \rho_{e-p} + \rho_{e-e}
\end{equation}
where $ \rho_{0} $ is the residual resistivity, $ \rho_{e-m} $,  $\rho_{e-p}$ and $\rho_{e-e}$ are the electron-magnon, electron-phonon and electron-electron scattering contributions to the total resistivity. $\rho_{0}$ is temperature independent part arising from lattice defects, imperfections or disorder. $\rho_{e-p}$ varies linearly with temperature whereas $\rho_{e-m}$ and $\rho_{e-e}$ have quadratic temperature dependence. The temperature dependent resistivity is analyzed in the temperature range of 5 K - 50 K, 50 K - 110 K and 150 K - 300 K in order to extract the different scattering contributions to the resistivity. Figure \ref{rtfitall} (a) shows the temperature dependent resistivity at zero-field along with the theoretical fits and \ref{rtfitall} (b) shows $\rho_{xx} (T)$ at various constant magnetic fields. In the high temperature range i.e. 200 K $\leq$ T $\leq$ 300 K , a linear temperature dependence of resistivity is observed as a result of the electron-phonon scattering. Below T$_{C}$, in the temperature range of 50 K $\leq$ T $\leq$ 110 K, a combination of linear and quadratic temperature dependence fits well. It is found that the linear term is an order of magnitude larger than the square term which suggests that the electron-phonon scattering is the dominant contribution in the temperature range upto 50 K. Below 50 K, only a T$ ^{2} $ dependence is obtained. These fittings have also been applied for the data in presence of magnetic fields but for the sake of clarity these fittings are not shown in the figure.
\begin{figure}[h!]
	\centering
	\includegraphics[width=\linewidth]{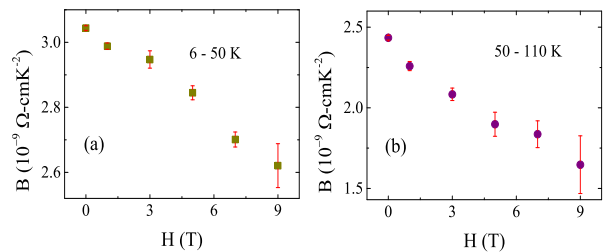}
	\caption{Variation in strength of electron-magnon scattering (B) with magnetic field in the temperature range (a) 6-50 K and (b) 50-110 K}
	\label{bcoeff}
\end{figure}
Scattering from both electron-electron and electron-magnon follows the T$ ^{2} $ relation but, the strength of scattering resulting from electron-electron is insensitive to magnetic fields whereas, the strength of electron-magnon scattering gets suppressed with external applied magnetic fields. To determine the origin of T$^{2}$ dependence, the magnetic field dependence of such scattering strength has been determined. Coefficient of T$^{2}$ term (B) obtained from the fitting between 5-50 K and 50-110 K are plotted against the respective fields in figures \ref{bcoeff} (a) and (b) showing a decreasing trend with the increase in magnetic field; suggesting the T$^2$ behaviour originates from electron-magnon scattering. The indication of electron-magnon scattering has also been reported by earlier specific heat and resistivity measurements \cite{graf2009,zhang2005}.
\begin{figure}[b!]
	\centering
	\includegraphics[width=0.7\linewidth]{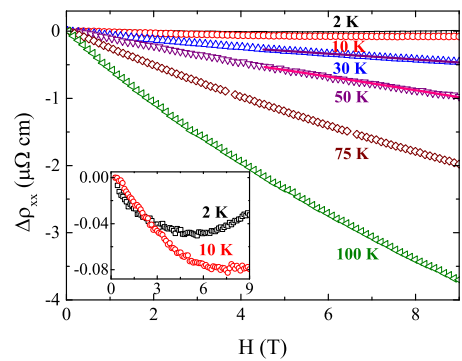}
	\caption{Change in resistivity ($ \Delta\rho_{xx}$) as a function of field measured at various constant temperatures, the solid lines are fit to the equation (\ref{eq:mr-equation}) at 30 K and 50 K. Inset shows the enlarged view of $\Delta\rho_{xx}$ at  10 K and 2 K.}
	\label{cta-mr}
\end{figure}
\par
Figure \ref{cta-mr} shows isothermal change in resistivity $\Delta\rho_{xx}(H)$= $\rho_{xx}(H)-\rho_{xx}(0)$ as a function of magnetic field below T$_C$. Resistivity decreases on increasing the field leading to negative magnetoresistance. The magnitude of change in resistivity decreases on lowering the temperature, and below 10 K, resistivity initially falls on increasing the field but shows an upturn at higher field. This is shown in inset of figure \ref{cta-mr}. At large fields, above the technical saturation, the field dependence of isothermal resistivity is mainly determined by electron-magnon scattering and scattering from Lorentz force. The spin flip scattering of conduction electrons with magnons gets suppressed on increasing the magnetic field which reduces the resistivity. On the other hand, the scattering from Lorentz contribution increases with magnetic field and enhances the resistivity. Below 100 K the non saturating decay of resistivity with field is an indication of the electron-magnon scattering. Magnons are the collective magnetic excitations of the spin ordered ground state, and at low temperatures, it becomes increasingly difficult to excite the magnons in the spin ordered state. This reduces the total amount of magnons and thus the electron-magnon scattering. The decrease in the magnitude of resistivity drop at lower temperatures is an outcome of reduced electron-magnon scattering. The high field dependence of resistivity due to electron-magnon scattering in the temperature range of T$_C$/5 to T$_C$/2, and below the fields of 100 T, can be described by \cite{raquetprb2002}:
\begin{equation}\label{eq:mr-equation}
	\Delta\rho_{xx} (T,B) \propto \frac{BT}{D(T)^2}ln(\frac{\mu_{B}B}{\kappa_{B}T}),
\end{equation}
where $D(T)$ is defined as magnon stiffness or magnon mass renormalization, $B$ is magnetic field and $T$ is the temperature. The fitting of equation (\ref{eq:mr-equation}) to isothermal resistivity curves at the temperatures of 50 K and 30 K above 4 T is shown in figure \ref{cta-mr}. The data fits well with equation (\ref{eq:mr-equation}) confirming that electron-magnon scattering is responsible for the observed non saturated negative magnetoresistance. Below 10 K, the strength of electron magnon scattering is weak which gets further suppressed on increasing the magnetic field.  At larger fields, the scattering from Lorentz contribution starts dominating leading to an upturn in magnetoresistance.

\subsection{Hall Resistivity}

\begin{figure}[h!]
	\centering
	\includegraphics[width=0.8\linewidth]{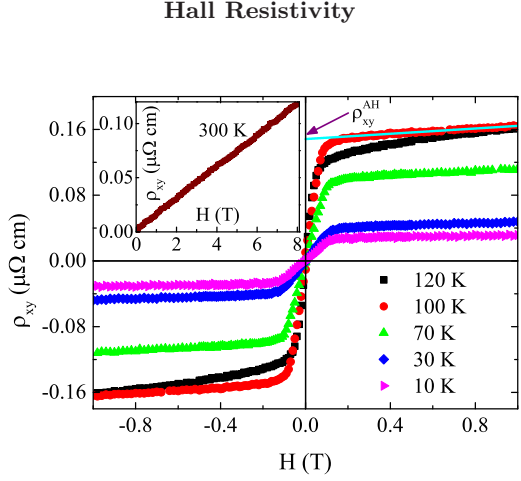}
	\caption{Isothermal Hall resistivity versus magnetic field at various temperatures.}
	\label{cta hall}
\end{figure}
\begin{figure}[h!]
	\centering
	\includegraphics[width=0.8\linewidth]{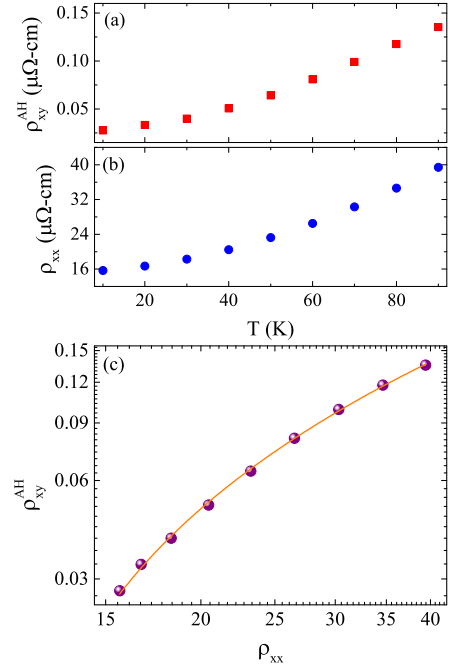}
	\caption{(a) $\rho_{xy}^{AH}$ and (b) $\rho_{xx}$ as a function of temperature. (c) shows the scaling between $\rho_{xy}^{AH}$ and $\rho_{xx}$ as per equation (\ref{modified scaling}).}
	\label{scaling}
\end{figure}

Hall resistivity as a function of magnetic field at constant temperatures below T$ _{C} $ is shown in figure \ref{cta hall}. Above T$_{C}$ in the PM state, hall resistivity increases linearly with magnetic field which is the ordinary hall effect as shown in the inset of figure \ref{cta hall}. Below T$ _{C} $ in the FM state, $\rho_{xy}$ has two linear regions; first is the low field region where it increases steeply up to a certain field above which there is a change in slope which almost saturates. This behaviour of hall resistivity in the FM state is the anomalous hall effect. The slope of $\rho_{xy}$ in the high field region renders the charge carrier density while the sign of slope determines the type of charge carriers. $\rho_{xy}^{AH}$ is determined by extrapolating the $\rho_{xy}$(H) data from high field region to zero field as shown in figure \ref{cta hall}. The temperature dependence of $\rho_{xy}^{AH}$ and $\rho_{xx}$ is shown in figure \ref{scaling} (a) and (b) respectively. It is observed that the anomalous hall resistivity increases with temperature similar to that of the longitudinal resistivity. The anomalous Hall resistivity ($ \rho_{xy}^{AH}$) and the longitudinal resistivity $\rho_{xx}$ follow a scaling relation based on the scattering mechanisms involved. In ferromagnets, the anomalous hall resistivity $\rho_{xy}^{AH}$ scales with longitudinal resistivity $\rho_{xx}$ as:
\begin{equation}\label{ahe scaling}
	\rho_{xy}^{AH} = a\rho_{xx} + b\rho_{xx}^2
\end{equation}
where the first and second terms on right hand side arise from skew scattering (SK) and side-jump (SJ) or intrinsic (I) contributions respectively to the anomalous hall resistivity \cite{smit1955,smit1958,berger1970prb,KL1954,luttinger1958}.
It has been shown that the temperature independent residual resistivity  ($\rho_{xx0}$) and temperature dependent part of the longitudinal resistivity ($\rho_{xxT}$) have distinct role in driving the AHE \cite{tian2009prl,hou2015prl}. Therefore, the scaling between the $\rho_{xy}^{AH}$ and $\rho_{xx}$ has been revised taking $\rho_{xx0}$ and $\rho_{xxT}$ both into account i.e. by replacing $\rho_{xx}$ by ($\rho_{xx0}$ + $\rho_{xxT}$), equation (\ref{ahe scaling}) can be written as:
\begin{equation}\label{modified scaling}
	\rho_{xy}^{AH} (T)= (\alpha_{0}\rho_{xx0} + \alpha_{1}\rho_{xxT}) + \beta_{0}\rho_{xx0}^2 + \gamma\rho_{xx0}\rho_{xxT} + \beta_{1}\rho_{xxT}^2
\end{equation}
Here the term ($\alpha_{0}\rho_{xx0} + \alpha_{1}\rho_{xxT}$) is the total skew scattering ($\rho_{xy}^{AH-SK}$) contribution to $\rho_{xy}^{AH}$ (T), $\gamma\rho_{xx0}\rho_{xxT}$ is the cross-term arising from competition between different scatterings and ($\beta_{0}\rho_{xx0}^2 + \beta_{1}\rho_{xxT}^2$) is the total contribution from the side-jump scattering and/or intrinsic effect ($\rho_{xy}^{AH-(SJ,I)}$). Equation (\ref{modified scaling}) depicts the temperature dependence of anomalous hall resistivity.
\begin{figure}[th!]
	\centering
	\includegraphics[width=0.9\linewidth]{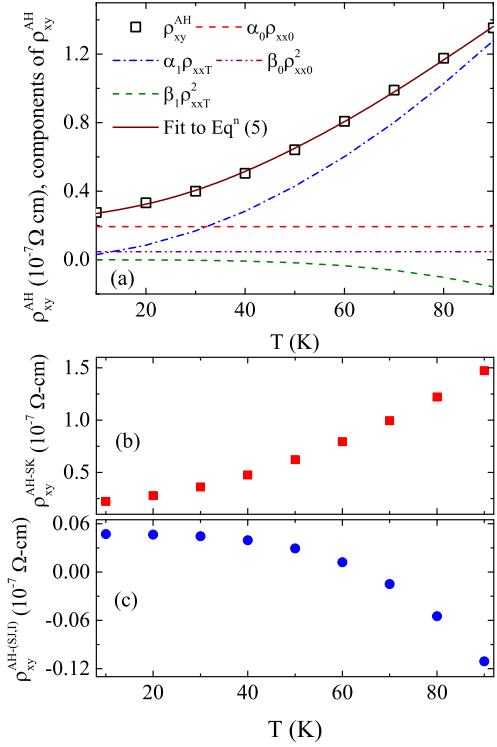}
	\caption{(a) Fitting of the equation (\ref{modified scaling}) to the temperature dependence of Anomalous hall resistivity ($\rho_{xy}^{AH}$) and the individual terms contributing to $\rho_{xy}^{AH}$ as a function of temperature, (b) and (c) show the dependence of Skew scattering and Side-jump/Intrinsic contributions to $\rho_{xy}^{AH}$ on temperature respectively.}
	\label{scaling fit}
\end{figure}
We have scaled $\rho_{xy}^{AH}$ with $\rho_{xx}$ in order to extract the scattering contributions that drive the AHE in this system. To begin, the general scaling for $\rho_{xy}^{AH}$ (T) defined in equation (\ref{ahe scaling}) is used to scale AHE but it does not provide a good fit to the data suggesting that it does not completely address the inherent mechanism involved in the AHE of this system. Individual linear and quadratic relation also does not provide the good fit. In order to separate the temperature dependent and temperature independent contributions to $\rho_{xy}^{AH}$, scaling of $\rho_{xy}^{AH}$ with $\rho_{xx}$ discussed in equation (\ref{modified scaling}) is used to fit the data and a good fit is obtained which is shown in figure \ref{scaling}(c). 
Figure \ref{scaling fit}(a) shows the dependency of each individual term used in equation (\ref{modified scaling}) on temperature, which are basically the different contributions to $\rho_{xy}^{AH}$ (T). The terms $\alpha_{0}\rho_{xx0}$ and $\beta_{0}\rho_{xx0}$ are temperature independent, whereas the terms $\alpha_{1}\rho_{xxT}$ and $\beta_{1}\rho_{xxT}^2$ are temperature dependent terms. The term $\beta_{1}\rho_{xxT}^2$ has a weak temperature dependence upto 50K and increases in magnitude above this temperature. 
\begin{figure}[h!]
	\centering
	\includegraphics[width=0.8\linewidth]{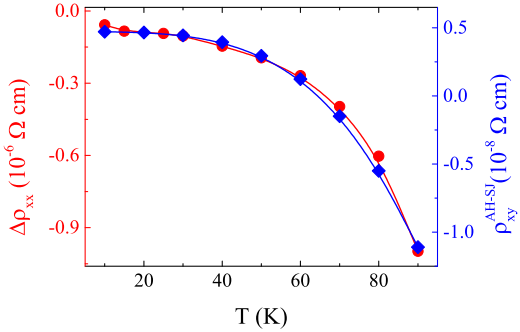}
	\caption{Change in resistivity ($\Delta\rho_{xx}$) (red solid circles) and the side-jump contribution to AHE ($\rho_{xy}^{AH-SJ}$) (blue solid rhombus) as a function of temperature (solid lines are the guide to eye).}
	\label{mr-ahe}
\end{figure}
Irrespective of magnitude, a finite contribution from all these terms is obtained at all the temperatures. Using the equation (\ref{modified scaling}), a good fit to the temperature dependence data of the $\rho_{xy}^{AH}$ (T) is obtained confirming the scaling of $\rho_{xy}^{AH}$ (T) with $\rho_{xx}$ as shown in figure \ref{scaling fit} (a). Figure \ref{scaling fit} (b) and (c) show the total contribution to the $\rho_{xy}^{AH}$ arising from skew scattering ($\rho_{xy}^{AH-SK}$) and side-jump or intrinsic effect ($\rho_{xy}^{AH-(SJ,I)}$) respectively where both make finite contribution to the total $\rho_{xy}^{AH}$ and are temperature dependent. In terms of magnitude, the skew scattering is almost an order larger than the side-jump/intrinsic contributions thereby dominating the $\rho_{xy}^{AH}$ (T).
\begin{figure}[h!]
	\centering
	\includegraphics[width=0.65\linewidth]{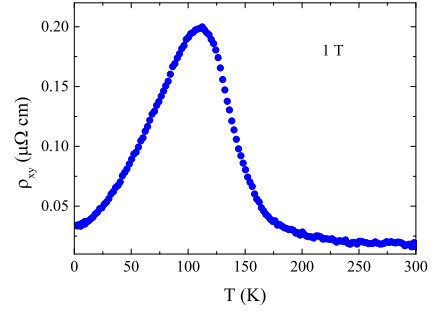}
	\caption{Hall resistivity as a function of temperature in the presence of 1 T magnetic field.}
	\label{hall-vs-t}
\end{figure}
Usually it is expected that the intrinsic contribution remains almost independent of temperature. Experimentally, it has been a challenge to distinguish the intrinsic and side-jump contributions as both show the scaling $\rho_{xy}^{AH} \propto \rho_{xx}^2$. Since the intrinsic contribution to AHE is considered to be temperature independent, the temperature dependent part of ($\rho_{xy}^{AH-(SJ,I)}$) is expected to be coming from side-jump scattering contribution  ($\rho_{xy}^{AH-SJ}$).

\par
Further it has been proposed that the electron-magnon scattering can lead to the temperature dependence of the side-jump contribution \cite{yang2011prb}. As discussed earlier, our system has electron-magnon contribution to resistivity and the magnetoresistance is governed by electron-magnon spin flip scattering. Therefore, magnons could be responsible for side-jump contribution asserting the theoretical prediction by Yang et al. \cite{yang2011prb}. To confirm this possibility, we have scaled the temperature dependence of change in resistivity with field ($\Delta\rho_{xx}$) and $\rho_{xy}^{AH-SJ}$ as shown in figure \ref{mr-ahe}. A good correlation between MR and $\rho_{xy}^{AH-SJ}$ confirms that the  $\rho_{xy}^{AH-SJ}$ (T) originates from the spin flip electron-magnon scattering.
\par
Figure \ref{hall-vs-t} shows hall resistivity as a function of temperature at 1 T. As temperature is reduced $\rho_{xy}$ slowly increases with a peak  at T$_{C}$ $\sim$ 125 K followed by reduction in $\rho_{xy}$. Large $\rho_{xy}$ at T$_{C}$ could be due to spin fluctuations where PM to FM order sets in. The temperature dependent $\rho_{xy}$ values are same as those obtained from isothermal $\rho_{xy}$ at 1 T which further supports the scaling of $\rho_{xy}^{AH}$ with $\rho_{xx}$ in the FM region.
\section{Summary}
In conclusion, a detailed study of magnetotransport properties of Cobalt-based Heusler alloy Co$ _{2} $TiAl by resistivity, magnetoresistance and hall resistivity has been carried out. Analysis of temperature dependent $\rho_{xx}$ shows the manifestation of electron-magnon scattering below T$_{C}$. In this regime the magnetoresistance is negative and governed by spin flip electron-magnon scattering. Hall resistivity shows a large change at PM to FM transition. Anomalous hall resistivity is observed below T$_{C}$ which scales with longitudinal resistivity. Scaling of anomalous hall resistivity shows that AHE in this system is driven by extrinsic mechanism viz. skew scattering and side-jump scattering mechanisms, however skew scattering dominates the $\rho_{xy}^{AH}$ (T). Side-jump contribution to the anomalous hall resistivity correlates well with the magnetoresistance, confirming the origin of side-jump contribution to be the electron-magnon scattering.

\section*{Acknowledgment}

Authors thank M. Gupta and L. Behera for XRD measurement.

\bibliography{cta}
\end{document}